\documentclass[aps,pra,twocolumn,showpacs]{revtex4}
\usepackage{amsmath}
\usepackage{dcolumn}
\usepackage{amssymb}
\usepackage{amsfonts}
\usepackage{txfonts}
\usepackage{latexsym}
\usepackage{euscript}
\usepackage{graphicx}
\usepackage[T1]{fontenc}
\usepackage{textcomp}
\begin{document}
\title{Feshbach resonant scattering of three fermions in one-dimensional wells}
\author{T. \DJ uri\'c$^{1}$, A. F. Ho$^2$, and  D. K. K. Lee$^{1}$}
\affiliation{$^1$Blackett Laboratory, Imperial College London, Prince Consort Rd, London SW7 2AZ, United Kingdom}
\affiliation{$^2$Department of Physics, Royal Holloway University of London, Egham, Surrey TW20 0EX, United Kingdom}
\date{\today}
\begin{abstract}
We study the weak-tunnelling limit for a system of cold $^{40}$K atoms
trapped in a one-dimensional optical lattice close to an $s$-wave
Feshbach resonance. We calculate the local spectrum for three atoms at
one site of the lattice within a two-channel model. Our results
indicate that, for this one-dimensional system, one- and two-channel models will differ close to the
Feshbach resonance, although the two theories would converge in the
limit of strong Feshbach coupling.  We also find level crossings in
the low-energy spectrum of a single well with three atoms that may
lead to quantum phase transition for an optical lattice of many wells.
We discuss the stability of the system to a phase with non-uniform
density.
\end{abstract}
\maketitle
\section{Introduction}
\label{sec:intro}
Quantum simulation of strongly correlated models of condensed matter systems has become one of  the most exciting goals in ultra-cold atom experiments, thanks to the spectacular progress in trapping cold fermions in optical lattices at degeneracy temperatures \cite{Bloch08}. This new system goes beyond traditional experiments in solids, because  it is readily possible to tune independently the hopping amplitude and interaction strength. In particular, by changing the external magnetic field that tunes the relative Zeeman energy between the scattering state and a closed (bound state) channel, one can reach  a particular Feshbach resonance between two hyperfine states of, for example, $^{40}$K where the scattering length between these states becomes very large and positive \cite{BCSBEC}. This then allows one to tune the on-site two-body interaction to be strongly repulsive. With this, the fermionic Mott insulator, where interaction driven locking of fermions into a crystalline structure, has been demonstrated in two recent experiments \cite{Jordens08, Schneider08}. Currently, there are great experimental interests to reach ordered phases in these systems, in order to understand the phase diagram of the two-dimensional single-band Hubbard model, a model that has been hypothesized to be the minimal model for the high temperature superconductors \cite{Ho08}.

Indeed, the unprecedented experimental possibility of studying extremely strong repulsion has spurred on much new theoretical effort \cite{Bloch08}. In particular, it was recognised that, close to a Feshbach resonance, the strongly enhanced two-body interaction leads to states with occupation of {\it multiple} Bloch bands of the optical lattice, both in experimental \cite{Kohl} and theoretical studies \cite{Busch, Diener}. These multi-band systems \cite{Duan,Ho} may lead to richer phases, by analogy with multi-orbital systems in solid state materials.

While the ultra-cold atoms parameters of the one-band Hubbard model have been derived right from the beginning \cite{Jaksch}, the parameters for the multi-band generalisation have only more recently received some attention \cite{Busch,Diener, Kestner, Mora, Werner, Castin}. Motivated by the experiments of K\"ohl {\it et al.} \cite{Kohl} where the average  filling per lattice site is about two fermions or slightly more, we study here the energy spectra when two or three fermions interact near a Feshbach resonance within a site of the deep optical lattice, using a two-channel description of the Feshbach resonance. This is then the first step towards deriving from microscopic physics, the effective model and its parameters for a system of strongly interacting fermions in an optical lattice.

In this paper, we consider a system of cold fermionic $^{40}$K atoms in the two lowest internal states trapped in a deep optical lattice close to an inter-species $s$-wave Feshbach resonance. We assume a lattice with average filling of two atoms per site with an equal mix of the two internal states. For simplicity, we consider only a one-dimensional (1D) system. In experiments with cold atoms in optical lattices, a set of quasi-1D tubes can be obtained from a 3D optical lattice, $V(\vec{r})=V_{0x}\sin^2k_xx+V_{0y}\sin^2k_yy+V_{0z}\sin^2k_zz$, by adjusting the laser amplitudes so that $V_{0z}\ll V_{0x}=V_{0y}$ \cite{Moritz,Moritz2}. We concentrate on the weak tunnelling limit (deep lattice), in which experiments with lattice fermions across a Feshbach resonance are usually performed \cite{Kohl}, and approximate the optical lattice potential at each lattice site by a harmonic well, $V_{ho}=\sum_{i=x,y,z}\mu\omega_i^2/2$ with $\mu\omega_i^2/2=2k_i^2V_{0i}$. The local spectrum and eigenstates for two and three atoms at a lattice site provide then a microscopic basis for the model of such lattices found by Ho \cite{Ho} where a weak tunnelling is introduced perturbatively. 

The spectrum for two atoms in a harmonic well close to a $s$-wave Feshbach resonance was previously studied in e.g. \cite{Busch} within the one-channel model for the harmonic confinement in one, two and three dimensions and in \cite{Diener} within the two-channel model for a three dimensional harmonic trap. However, the spectrum for three atoms in a well was previously studied only within the one-channel model (e.g. for 3D harmonic confinement\cite{Kestner} , quasi-1D confinement\cite{Mora} and fermions at unitarity\cite{Werner,Castin}). Here, we calculate the spectrum within the two-channel model and examine the ground state for our system across the resonance. We find the level crossings in the low energy spectrum that may lead to quantum phase transition when multiple wells form an optical lattice. Similar features were found in the spectrum for a 3D system in \cite{Kestner} within the one-channel model. However, we find that there are quantitative differences between our two-channel and their one-channel results at and near the Feshbach resonance.

\section{Two-channel model for three atoms in a well}
\label{sec:model}
When $V_{0z}\ll V_{0x}=V_{0y}$, the system can be described by an effective 1D Hamiltonian. The effective 1D parameters can be related to the physical 3D parameters in the limit $\omega_z\rightarrow 0$ as presented in \cite{Yurovski}. We concentrate on the weak tunnelling limit and calculate the local spectrum when there are three atoms in two different internal states in the well. 

The $^{40}$K atoms are assumed to be trapped in the two lowest
hyperfine energy states $|F_1,m_{F_1}\rangle=|9/2,-9/2\rangle$ and
$|F_2,m_{F_2}\rangle=|9/2,-7/2\rangle$. The interatomic interaction
potential,
$U(\vec{r}_1-\vec{r}_2)=U_c(\vec{r}_1-\vec{r}_2)+U_s(\vec{r}_1-\vec{r}_2)\vec{S}_1\cdot\vec{S}_2$,
depends only on electronic spins of atoms and preserves the total
angular momentum projection of the two colliding atoms,
$m_F=m_{F_1}+m_{F_2}$. Then, for the two $^{40}$K atoms in the two
lowest internal states that constitute the open entrance channel,
$|\alpha_1,\alpha_2\rangle=|9/2,-9/2;9/2,-7/2\rangle$, the only closed
channel coupled to this state is $|9/2,-9/2;7/2,-7/2\rangle$. For
simplicity, we will use here the notation $|9/2,-9/2\rangle \equiv
|\uparrow\rangle$, $|9/2,-7/2\rangle \equiv |\downarrow\rangle$ and
$|7/2,-7/2\rangle \equiv |\circ\rangle$. In this notation, the
Feshbach physics relies on the scattering between the states:
\begin{equation}\label{eq:feshbachscatter}
 |\uparrow \downarrow\rangle \quad \leftrightarrow\quad  |\uparrow\circ\rangle
\end{equation}
where the final closed-channel state is a tightly-bound $|\uparrow\circ\rangle$ dimer.
We are interested in a regime near a Feshbach resonance so that the Feshbach-induced interaction is strong. 
We will therefore neglect for simplicity the direct $s$-wave scattering of the open-channel states. 

In this section, we will focus on the case where there are two atoms
of the species $\uparrow$ and one atom of the species $\downarrow$ in
the well. (We will comment on the case of one $\uparrow$ and two
$\downarrow$ atoms at the end of the section.) We will calculate how
the open-channel levels $|\uparrow,\uparrow,\downarrow\rangle$
(\emph{i.e.}, atoms 1 and 2 are $\uparrow$ and atom 3 is $\downarrow$)
become coupled together with levels involving a pair of atoms in the
closed channel $|\uparrow,\uparrow,\circ\rangle$.

Consider first the states in the open channel in the absence of Feshbach scattering.
Since we are ignoring direct $s$-wave scattering in the open channel, 
atoms in the open-channel states are non-interacting and they will have a Hamiltonian 
$h^{\rm op}$ that consists only of the potential of the harmonic trap $h^{\rm trap}$.

We will work in the centre-of-mass frame of this three-body system
because the motion of the centre of mass can be factored out
completely for a single harmonic well.  We denote the positions of the atoms
$\uparrow,\uparrow,\downarrow$ with $\vec{r}_1=r_1\hat{z}$,
$\vec{r}_2=r_2\hat{z}$ and $\vec{r}_3=r_3\hat{z}$, respectively.  It
is conventional to describe the relative degrees of freedom of the
system using the Jacobi coordinates:
$x_1=2\left[r_3-(r_1+r_2)/2\right]/\sqrt{3}$ and $x_2=r_2-r_1$. We
find that it is more convenient for the evaluation of our formulae to
use a rotated version of these coordinates:
\begin{eqnarray}\label{eq:coords}
x &=& \frac{x_1-x_2}{\sqrt{2}} = \frac{1}{\sqrt{6}}[2r_3+(\sqrt{3}-1)r_1 - (\sqrt{3}+1)r_2]\nonumber\\
y &=& \frac{x_1+x_2}{\sqrt{2}} = \frac{1}{\sqrt{6}}[2r_3-(\sqrt{3}+1)r_1 + (\sqrt{3}-1)r_2]
\end{eqnarray}
The potential of the harmonic trap is separable in these coordinates:
\begin{equation}\label{eq:hamopen}
h^{\rm op} = h^{\rm trap}(x,y)=-\frac{\hbar^2}{2\mu}\left(\nabla_x^2+\nabla_y^2\right)+\frac{\mu\omega^2}{2}\left(x^2+y^2\right)
\end{equation}
where $\mu=m/2$ is the reduced mass for the relative motion of two atoms. 
The characteristic length of the oscillations in the harmonic trap is $d_r = (\hbar/\mu\omega)^{1/2}$ in the centre-of-mass frame. The eigenstates of this Hamiltonian must have wavefunctions which are antisymmetric under the exchange of two fermions in the same internal state 
($\uparrow$), \emph{i.e.} under the exchange $r_1\leftrightarrow r_2$ which corresponds to $x\leftrightarrow y$. Therefore, the eigenstates are 
\begin{equation}\label{eq:basisopen}
|k,l\rangle=   \frac{\left[u_k(x,d_r)u_l(y,d_r)-u_k(y,d_r)u_l(x,d_r)\right]}{\sqrt{2}}
\,|\uparrow,\uparrow,\downarrow\rangle
\end{equation}
for non-negative integers $k\neq l$ and
\begin{equation}\label{eq:hermite}
u_p(s,d)=\frac{e^{-s^2/2d^2}}{(\pi d^2)^{1/4}(2^pp!)^{1/2}}\cdot H_p\left(\frac{s}{d}\right)
\end{equation}
with $H_p$ being the Hermite polynomials of order $p=0,1,\ldots$. 
These states have energy $(k+l+1)\hbar\omega$. Note that Pauli exclusion is enforced in that the wavefunction vanishes unless $k\neq l$.

Having found basis states for the open channel, we can do the same for
the closed channel with atoms 1 and 2 in the $\uparrow$ state and atom
3 in the $\circ$ state. They will have an energy relative to the open
channel of $\delta_B$ which is tunable by an external magnetic field
$B$. Moreover, there is an additional attractive
interaction $V^{\rm cl}$ in the closed channel that creates the
$|\uparrow\circ\rangle$ dimer essential for Feshbach physics.  
Therefore, the closed-channel Hamiltonian $h^{\rm cl}$ is given by:
\begin{equation}\label{eq:hamclosed}
h^{\rm cl}=h^{\rm trap}+V^{\rm cl}(r_{+})+V^{\rm cl}(r_{-})+\delta_B
\end{equation}
where the coordinates $r_{\pm}=r_3-r_{1,2}$
represent distances between atom 3 (in $\circ$ state in the closed
channel) and each of the other two $\uparrow$ atoms.

To find the eigenstates of this closed-channel Hamiltonian $h^{\rm
cl}$, it is useful to identify parts of $h^{\rm cl}$ which provide an attractive
interaction between one pair of $\uparrow\circ$ atoms to form a dimer
while leaving the other $\uparrow$ atom as a spectator.
Let $h_+^{\rm cl}$ provide an attraction between atoms 1 and 3, leaving atom 2 as a spectator
and $h_-^{\rm cl}$ provide an attraction between atoms 2 and 3, leaving atom 1 as a spectator.

These truncated dimer-spectator Hamiltonians are:
\begin{equation}\label{eq:hamclosedspectator}
h_{\pm}^{\rm cl}=-\frac{\hbar^2}{2\mu}\left(\nabla^2_{s_{\pm}}+\nabla_{r_{\pm}}^2\right)+
\frac{\mu\omega^2}{2}\left(s_{\pm}^2+r_{\pm}^2\right)+V^{\rm cl}(r_{\pm})+\delta_B
\end{equation}
where $s_\pm= [r_{1,2}+r_3 - 2r_{2,1}]/\sqrt{3}$
are proportional to the distances between the centre of mass of the dimer and the spectator. 

The eigenstates of these truncated Hamiltonians
(\ref{eq:hamclosedspectator}) are atom-dimer states of the form
$u_n(s_\pm,d_r)\chi(r_{\pm},d_m)$. The first part, $u_n$, represents the
oscillation in the relative displacement between the spectator and the dimer
with energy $(n+1/2)\hbar\omega+\delta_B$.
The second part, $\chi(r_{\pm},d_m)$, is the wavefunction for
the bound state formed due to the attractive interaction $V^{\rm cl}$:
\begin{equation}\label{eq:boundstate}
\left[-\frac{\hbar^2}{2\mu}\nabla_{r_{\pm}}^2+
\frac{\mu\omega^2}{2}r_{\pm}^2+V^{\rm cl}(r_{\pm})\right]\chi(r_{\pm})=\epsilon_b\chi(r_{\pm})
\end{equation}
with energy $\epsilon_b$. We cannot solve the bound-state equation in
general. We will simply assume that $\chi$ is a tightly bound $s$-wave
state much smaller in size than the characteristic trap oscillation amplitude
$d_r$: $\chi(r_{\pm},d_m)\simeq u_0(r_{\pm},d_m)$ with $d_m\ll d_r$.

However, the eigenstates of $h_{\pm}^{\rm cl}$ are not exact
eigenstates of the total closed channel because they do not include
the interaction of the spectator atom with the dimer in the state
$\chi$. Nevertheless, in the regime of a tightly bound dimer, 
this residual atom-dimer interaction is expected to be small due to Pauli suppression.
Suppose that atoms 1
and 3 have formed a dimer and atom 2 is the $\uparrow$ spectator. The
$u_n(s_+,d_r)$ part of the wavefunctions indicates that atoms
1 and 2 are separated at a distance of the order of
$(3n)^{1/2}d_r/2$.  Moreover, since they are fermions, Pauli
exclusion means that their relative wavefunction must have a node at
$r_1=r_3$, so that $u_n$ remains small until they are separated by a
length scale set by the trap length $d_r$. Since atom 3 stays close to
atom 1 in a tight dimer, this means that atoms 2 and 3 also stay apart
at a distance of the order of $d_r$ where the interaction $V^{\rm cl}$ between them
is weak. Therefore, we can, to a first approximation, neglect this
spectator-dimer interaction so that
approximate eigenstates for the closed-channel Hamiltonian $h^{\rm
cl}$ are antisymmetrised combinations of the two possible
spectator-dimer states:
\begin{equation}\label{eq:basisclosed}
|\chi_n\rangle=  \frac{\left[
u_n(s_+,d_r)\chi(r_+,d_m)-u_n(s_-,d_r)\chi(r_-,d_m)\right]}{\sqrt 2}
\,|\uparrow,\uparrow,\circ\rangle
\end{equation}
for $n=0,1,\ldots$, with energies $(n+1/2)\hbar\omega + \epsilon_b + \delta_B$.

We can now discuss the full Hamiltonian with Feshbach scattering between the open and closed channels.
The Hamiltonian for the three atoms is given (in the centre-of-mass frame) by:
\begin{eqnarray}\label{eq:hamiltonian}
\hat{H}^{\uparrow\uparrow\downarrow}
&=&
\left[\begin{array}{cc}|\uparrow,\uparrow,\downarrow\rangle&|\uparrow,\uparrow,\circ\rangle\end{array}\right]\left[\begin{array}{cc}h^{\rm op}&W\\W&h^{\rm cl}\end{array}\right]\left[\begin{array}{c}\langle\uparrow,\uparrow,\downarrow|\\\langle\uparrow,\uparrow,\circ|\end{array}\right]\nonumber\\
W&=&\alpha(r_+)+\alpha(r_{-})
\end{eqnarray} 
where $\alpha(r)$ describes the coupling between the open and closed channels
for two atoms at a distance of $r$. 
The eigenstates of this Hamiltonian (\ref{eq:hamiltonian}) are now superpositions of the 
open and closed channel states:
\begin{equation}\label{eq:eigenstate}
|\psi^{\uparrow\uparrow\downarrow}\rangle=\sum_{kl\atop k<l}a_{kl}|kl\rangle+\sum_nb_n|\chi_n\rangle
\end{equation}
where the first term represents the three atoms in the open channel in terms of the basis set (\ref{eq:basisopen}) 
and the second term represents the atom-dimer states in terms of the basis set (\ref{eq:basisclosed}).
An eigenstate with energy $E$ obeys the Schr\"odinger equation:
$H^{\uparrow\uparrow\downarrow}|\psi^{\uparrow\uparrow\downarrow}\rangle=
E|\psi^{\uparrow\uparrow\downarrow}\rangle$.
This requires
\begin{eqnarray}\label{eq:seculareqn}
\frac{1}{2}\sum_{kl}\alpha_{kln}a_{kl}+b_n(\varepsilon_n+\bar{\nu})&=&Eb_n,\nonumber\\
\sum_n\alpha_{kln}b_n+(\varepsilon_{k}+\varepsilon_{l})a_{kl}&=&Ea_{kl},
\end{eqnarray}
where $\varepsilon_n=(n+1/2)\hbar\omega$, $\alpha_{kln}=\langle
kl|W|\chi_n\rangle$ is the Feshbach scattering matrix element, and $\bar{\nu} = \delta_B + \epsilon_b$ is a
detuning parameter which sweeps across the Feshbach resonance as a
function of applied magnetic field.

In the absence of the confinement potential, the resonance occurs when
the atom-dimer state of energy $\epsilon_b+\delta_B$ coincides in energy with the
zero-momentum scattering state (energy zero) in the open channel.
This corresponds to $\bar{\nu}=0$. In the presence of the harmonic
trap, the Feshbach scattering may cause resonances whenever
the open and closed channel states are degenerate:
\begin{equation}\label{eq:crossing}
(k+l+1)\hbar\omega = \bar\nu + (n+1/2)\hbar\omega
\end{equation}
for non-negative integers $k,l(\neq k)$ and $n$. A non-zero Feshbach coupling will lift the degeneracy
and one will find a level anti-crossing in the energy spectrum as a function of $\nu$.
We will see in section \ref{sec:selectionrules} that there are level crossings allowed by selection rules.

To simplify our calculation, we can eliminate the open-channel
coefficients $a_{kl}$ using the second
equation in (\ref{eq:seculareqn}). Then, we
obtain a matrix equation for just the atom-dimer components, $b_n$, of
the eigenstate:
\begin{equation}\label{eq:coeffatomdimer}
(E-\varepsilon_n) b_n + \sum_mM_{nm}(E)b_m=\bar{\nu}b_n,
\end{equation}
where
\begin{equation}\label{eq:Mmatrix} 
M_{nm}(E)=\frac{1}{2}\sum_{kl}
\frac{\alpha_{kln}\alpha_{klm}}{\varepsilon_k+\varepsilon_l-E}.
\end{equation}
The matrix $M_{nm}$ can be interpreted as representing an effective scattering of the closed-channel
state $|\chi_m\rangle$ to $|\chi_n\rangle$ \emph{via} the virtual state $|kl\rangle$ in the open channel.
We solve equation (\ref{eq:coeffatomdimer})
numerically by treating it, at any given $E$, as a matrix eigenvalue problem for eigenvalue $\bar\nu$.

We need to evaluate the Feshbach matrix elements $\alpha_{kln}$. 
We will assume that it only occurs when two atoms are
close together at a scale much smaller than the trap length
$d_r$. We see that there are direct and exchange terms in this matrix element.
The direct term involves  $\alpha(r_\pm)\chi(r_\pm,d_m)$.
For a tightly-bound dimer in the closed channel, we can reduce this to an effective delta-function coupling
between the open and closed channel\cite{Diener,Duine}: $\alpha(r)\chi(r,d_m)\rightarrow \alpha
\delta(r)$. This leaves $\alpha$ as a single parameter controlling the
strength of the Feshbach interaction. There is also an exchange term
involving $\alpha(r_{\mp})u_n(s_{\pm},d_r)\chi(r_\pm,d_m)$. This term is only large when 
all three atoms are very close to each other.
These terms can be neglected due to Pauli suppression by the same argument that
we neglected atom-dimer interactions in the closed channel: 
the range of the Feshbach scattering is much smaller that the typical separation ($\sim d_r$)
of two atoms in the same hyperfine state. 

With this approximation of a short-ranged Feshbach interaction, the
matrix elements $M_{nm}(E)$ can be written in terms of the
Green's functions of the system:
\begin{eqnarray}\label{eq:ME1}
M_{nm}(E)&=&\frac{\alpha^2}{4}\!\!\int \!\!dx dx' dy dy' \;
\psi_n(x,y)\psi_m(x',y')\\
&\times&\left[G^{(2)}(x,y;x',y';E)-G^{(2)}(x,y;y',x';E)\right]\nonumber
\end{eqnarray}
where $\psi_n=u_n(s_{+},d_r)\delta(r_+)-
u_n(s_{-},d_r)\delta(r_-)$ and $G^{(2)}$ is the two-particle
Green's function of non-interacting atoms in a 1D harmonic well
\begin{equation}\label{eq:ME2}
G^{(2)}(x,y;x',y';E)=\sum_{kl}\frac{u_k(x,d_r)u_l(y,d_r)u_k(x',d_r)u_l(y',d_r)}{\varepsilon_k+\varepsilon_l-E}.
\end{equation}
We notice \cite{Kestner} that the harmonic oscillator Hamiltonian 
$h^{\rm trap}$ is invariant under rotations in the $x$-$y$ plane
and that the ($r_\pm$,$s_\pm$) are both related to ($x,y$) by such a rotation:
$r_\pm = [(\sqrt{3}\mp 1)x + (\sqrt{3}\pm 1)y]/2\sqrt{2}$ and 
$s_\pm = [(1\pm \sqrt{3})x + (1\mp \sqrt{3})y]/2\sqrt{2}$.
Therefore, $G^{(2)}(x,y;x'y';E) = G^{(2)}(r_\pm,s_\pm;r_\pm',s_\pm';E)$.
This allows us to express all the integrations in terms of $r$ and $s$ variables.
We find that the matrix elements $M_{nm}(E)$ can be written in the final form 
\begin{eqnarray}\label{eq:ME3}
&
&M_{nm}(E)=\alpha^2\Bigg\{G^{(1)}(0,0;E-\varepsilon_n)\,\delta_{n,m}\\
&-&\int\!\!
ds\;G^{(1)}\left(\frac{\sqrt{3}s}{2},0;E-\varepsilon_n\right)u_n\left(-\frac{s}{2},d_r\right)u_m(s,d_r)\Bigg\},\nonumber
\end{eqnarray}
where $G^{(1)}$ is the single-particle Green's function
$G^{(1)}(x,x';E)=\sum_nu_n(x,d_r)u_n(x',d_r)/(\varepsilon_n-E)$.  The
advantage of expressing the matrix in this form is that we know the
quantities we need analytically \cite{Bakhrakh}:
\begin{equation}\label{eq:ME4}
G^{(1)}(x,0;E)=\frac{\Gamma\left(-\frac{E}{\hbar\omega}+\frac{1}{2}\right)}{\sqrt{\pi}\, d_r\, 
\hbar\omega}\, D_{\frac{E}{\hbar\omega}-\frac{1}{2}}\left(\frac{\sqrt{2}\;|x|}{d_r}\right)\, D_{\frac{E}{\hbar\omega}-\frac{1}{2}}(0)\,,
\end{equation}
where $D_{\nu}(x)$ is the parabolic cylinder function with
$D_{\nu}(|y|)=2^{\nu/2}e^{-y^2/4} U(-\nu/2,1/2,y^2/2)$ where $U$ is
the confluent hypergeometric function and
$D_{\nu}(0)=2^{\nu/2}\sqrt{\pi}/\Gamma\left[(1-\nu)/2\right]$.

The delta functions in the $\psi_n\psi_m$ factors in the form
(\ref{eq:ME1}) indicate which particles are interacting via the
Feshbach scattering.  The two $G^{(2)}$ terms comes from the fact the
virtual excitation \emph{via} an open-channel state involves a direct
process and an exchange process.  The direct process gives the first
term in (\ref{eq:ME3}) which requires $r_+=r_3-r_1=0$ or
$r_-=r_3-r_2=0$: it involves only the interaction of one pair of atoms
while the third atom remains a spectator. The exchange process
involves the virtual dissociation of a dimer into $\uparrow$ and
$\downarrow$ atoms in open-channel followed by the $\downarrow$ atom
recombining with a different $\uparrow$ atom to form a new dimer.
This gives the second term in (\ref{eq:ME3}) which is a term present
in the three-body problem but not in a two-body problem.

Before proceeding to discuss our results, we should point out the
difference between $|\uparrow,\uparrow,\downarrow\rangle$ and 
$|\downarrow,\downarrow,\uparrow\rangle$ states. Since the
Feshbach interaction only affects the $\downarrow$ state in the open
channel, there is no symmetry under the operation
$\uparrow\leftrightarrow\downarrow$.  The difference is that, in the
$\downarrow\downarrow\uparrow$ case, there is no residual interaction
between the $|\circ\rangle$ atom in the dimer and the spectator
$\downarrow$ atom:  the closed-channel attraction $V^{\rm
cl}$ only acts on a $|\uparrow\circ\rangle$ pair. However, we have already dropped this interaction within our
approximation of a tightly bound dimer state. Therefore, our results
should apply also to the $\downarrow\downarrow\uparrow$ case.  More
explicitly, for the $\downarrow\downarrow\uparrow$ case, the
Hamiltonian for the relative degrees of freedom is of the form
\begin{equation}\label{eq:hamdowndownup}
H^{\downarrow\downarrow\uparrow}=\textbf{V}\cdot
\left[\begin{array}{ccc}h^{\rm op}&W_+&W_-\\W_+&h^{\rm cl}_+&0\\W_-&0&h^{\rm cl}_-\end{array}\right]\cdot\textbf{V}^{\dagger}.
\end{equation}
where $W_{\pm}=\alpha(r_{\pm})$,   $\textbf{V}=[\begin{array}{ccc}|\downarrow,\downarrow,\uparrow\rangle&
|\circ,\downarrow,\uparrow\rangle&|\downarrow,\circ,\uparrow\rangle\end{array}
]$ and and $h^{\rm cl}_\pm$ have been defined in (\ref{eq:hamclosedspectator}). Now, the wavefunctions
\begin{equation}\label{eq:basisclosed2}
u_n(s_{+},d_r)\chi(r_+,d_m)|\circ,\downarrow,\uparrow\rangle
-u_n(s_{-},d_r)\chi(r_-,d_m)|\downarrow,\circ,\uparrow\rangle
\end{equation}
are exact eigenstates of the closed-channel part of the Hamiltonian.

\section{Selection Rules}
\label{sec:selectionrules}

We can take advantage of selection rules for the Feshbach coupling. 
In the pseudospin language of $\uparrow$ and $\downarrow$ states of the open
channel, we are discussing the three-atom states with $S_z=+1/2$. If
we completely antisymmetrise the wavefunction of the three atoms
\cite{Baym}:
\begin{equation}\label{eq:totalantisym}
|\psi(1,2,3)\rangle_A=|\psi(1,2,3)\rangle+|\psi(2,3,1)\rangle+|\psi(3,1,2)\rangle,
\end{equation}
we can classify the three-atom states as $S=3/2$ or $1/2$. The $S=3/2$
spin wavefunction is totally symmetric under exchange and so its
spatial part is totally antisymmetric under exchange, leading to a
node at $r_3=r_{1,2}$. Therefore, this state is unaffected by the
$s$-wave Feshbach scattering. On the other hand, the $S=1/2$ states
have mixed exchange symmetry (see for example \cite{Flugge}) and
Feshbach scattering is allowed.  To see this explicitly in our
formulation of the problem, we can write the open-channel part of the
Hamiltonian in terms of the polar coordinates $\zeta=(x^2+y^2)^{1/2}$
and $\phi=\arctan(y/x)$. In this coordinate system, the energies can
be written as $(2n_r+m_l+1)\hbar\omega$ with integers $n_r\geq 0$ and
$m_l > 0$.  The $S=3/2$ states correspond to $m_l=0$ (mod 3) and these
states have zero scattering matrix element to the atom-dimer
states. The $S=3/2$ state of lowest energy has energy $4\hbar\omega$.

In addition to exchange symmetry, we note that the system is symmetric
under inversion ($r_{1,2,3}\rightarrow -r_{1,2,3}$). This means that
the eigenstates of the system must be even or odd under inversion.  In
terms of the matrix equation (\ref{eq:coeffatomdimer}), the solutions
divide into two sectors, one for $b_n$ with even $n$ and one for $b_n$
with odd $n$. More precisely, the matrix elements $\alpha_{kln}$ are
non-zero only if $k+l$ and $n$ are both even or both odd.

There are also other matrix elements $\alpha_{kln}$ which vanish
between the open and closed channels: it can be shown that
$\alpha_{kln}$ vanish when $n> k+l$. This can be proved using the fact
that $\int_{-\infty}^{\infty}x^re^{-x^2}H_n(x)dx =0$ for $r<n$
\cite{Gradshteyn}. In other words, atom-dimer states with energy
$\bar{\nu}+(n+1/2)\hbar\omega$ do not couple directly to the
three-atom states with the relative energy $(k+l+1)\hbar\omega$ when
$n>k+l$, even if they are degenerate (possible when
$\bar\nu<\hbar\omega/2$). However, they can be coupled by
\emph{multiple} Feshbach scattering. Therefore, we expect avoided
level crossings for such levels but the splitting at the anti-crossing
point will be smaller in size than the level splittings at
anti-crossing points for levels coupled directly by a non-zero matrix
element $\alpha_{kln}$.  This is because the splitting will be higher
order in $\alpha$.

\section{Two-atom system}
\label{sec:twobody}
In this section, we will outline the results for the Feshbach resonances of two atoms in a 1D harmonic well.
This will be useful for the discussion of our results for three atoms.
These results are analogous to the 3D results of Diener and Ho \cite{Diener}. Our calculation differs from \cite{Diener} in the lower dimensionality and also in the fact that we have factored out the motion of the centre of mass.

In the centre-of-mass frame, the energy for two Feshbach interacting atoms in a well within the two-channel model is given by the equation
\begin{equation}\label{eq:twobodyenergy}
E-\bar{\nu}=-\frac{\alpha^2}{2 \hbar\omega d_r}\cdot\frac{\Gamma\left(\frac{1}{4}-\frac{E}{2\hbar\omega}\right)}{\Gamma\left(\frac{3}{4}-\frac{E}{2\hbar\omega}\right)}.
\end{equation}
The lowest-energy solution of this equation has an energy below $\hbar\omega/2$ for any value of the detuning parameter $\bar{\nu}$. Since $\hbar\omega/2$ is the lowest energy of the open-channel scattering states, this corresponds to a bound state. The existence of a bound state for any $\bar{\nu}$ is not unexpected in one dimension.

We can relate our two-channel parameters $\alpha$ and $\bar{\nu}$ to the parameters of a one-channel scattering theory in the absence of the trap: $a_{\rm 1D}$,
the $s$-wave scattering length, and $r_{\rm 1D}$, the effective range. These are the physical parameters that can be found experimentally. To do so, we follow the procedure of Diener and Ho \cite{Diener} and examine the bound state energy of two atoms in the limit of an infinitely shallow well:  $\omega\rightarrow 0$ at a constant negative $E$. We also need $|E|\ll\bar{\nu}$ so that the state is clearly separated from dimer states (energy $\bar\nu$) in the closed channel. In this limit, the equation (\ref{eq:twobodyenergy}) becomes
\begin{equation}\label{eq:twobodyenergy2}
\frac{\mu\alpha^2}{\hbar^2\bar{\nu}}-
\frac{\mu\alpha^2|E|}{\hbar^2\bar{\nu}^2}-\left(\frac{2\mu|E|}{\hbar^2}\right)^{1/2}=0.
\end{equation}
On the other hand, we find that the energy of a weakly bound state in a homogeneous case with no external confinement is given by the equation $1/a_{\rm 1D}-\mu r_{\rm 1D}|E|/\hbar^2-(2\mu|E|/\hbar^2)^{1/2}=0$ where $|E|=\hbar^2\kappa^2/(2\mu)$ and $\kappa R\ll 1$ with $R$ being the range of the scattering potential. 

Comparing those two equations, we find
\begin{eqnarray}\label{eq:scatteringparms}
1/a_{\rm 1D}&=&\mu\alpha^2/(\hbar^2\bar{\nu})\nonumber\\
r_{\rm 1D}&=&\alpha^2/\bar{\nu}^2.
\end{eqnarray}
We see that the effective 1D scattering length $a_{\rm 1D}$ is proportional to the detuning parameter $\bar\nu$
in the regime of a weakly-bound state where $|E| \ll \bar\nu$. In terms of the parameters of the two-channel model, 
this regime corresponds to a detuning of $\bar\nu \gg (\mu\alpha^4/2\hbar^2)^{1/3}$.
Note that this 1D case is very different to the 3D case where $a_{3D} \sim -1/{\bar \nu}$, see section \ref{sec:onechannel}.

\section{Results for three atoms}
\label{sec:threeatoms}
We will now discuss our results for the spectrum and eigenstates of the three-body system. 
As mentioned already, we treat  the matrix equation (\ref{eq:coeffatomdimer}) as an eigenvalue problem for the detuning parameter $\bar\nu$ at fixed energies $E$.

In order to obtain numerical results, we truncate the matrix equation (\ref{eq:coeffatomdimer}) by taking into account only atom-dimer energy levels with $\varepsilon_n<\varepsilon_{2n_C}$. This should properly reproduce the 
low-energy behaviour of the system for the few lowest atom-dimer energy levels. Here we choose $n_C=20$. We have checked that the ground and first excited states do not change significantly if we use a higher cutoff $n_C$.

We will discuss two opposite regimes of weak and strong Feshbach scattering.
The kinetic energy of the system is on the energy scale $\hbar\omega$. 
This should be compared to the magnitude of the scattering matrix elements $\alpha_{kln}$.
We note that  the oscillator wavefunctions $u_n(x,d_r)$ have a magnitude of the order of 
$d^{-1/2}_r$ over a spatial size $d_r$ so that $\alpha_{kln} \sim \alpha/d_r^{1/2}$. 
Therefore, a dimensionless measure for the strength of the Feshbach scattering is the ratio of these two energy scales:
\begin{equation}\label{eq:dimless}
\tilde{\alpha}=\frac{\alpha}{\hbar\omega d_r^{1/2}} \;.
\end{equation}
We will consider first the weak-scattering regime, $\tilde\alpha \ll 1$, where the open and closed channel states are weakly coupled. Then, we will discuss strong scattering, $\tilde\alpha \gg 1$, relevant to $^{40}$K atoms \cite{Yurovski,Kestner2}.

We can also obtain a scale for the range of detuning $\delta\bar\nu$ over which the Feshbach scattering significantly affects the atom-dimer energy levels, as governed by (\ref{eq:coeffatomdimer}). 
This can be estimated by the scale of the elements of $M_{nm}\sim (\alpha/d_r^{1/2})^2/\hbar\omega$.
Therefore, a dimensionless measure of the detuning parameter is $\bar\nu\hbar\omega d_r/\alpha^2$.
From the results (\ref{eq:scatteringparms}) for the two-body problem, we see that this is also the ratio of the effective 1D scattering length, $a_{\rm 1D}$, for states in the open channel (in the absence of a trap) compared to the trap oscillation length $d_r$:
\begin{equation}\label{eq:a1D}
\frac{a_{\rm 1D}}{d_r} = \frac{\bar\nu\hbar\omega d_r}{\alpha^2} 
\end{equation}
in the regime of a weak two-body bound state.

\begin{figure}[hbt]
\includegraphics[width=\columnwidth]{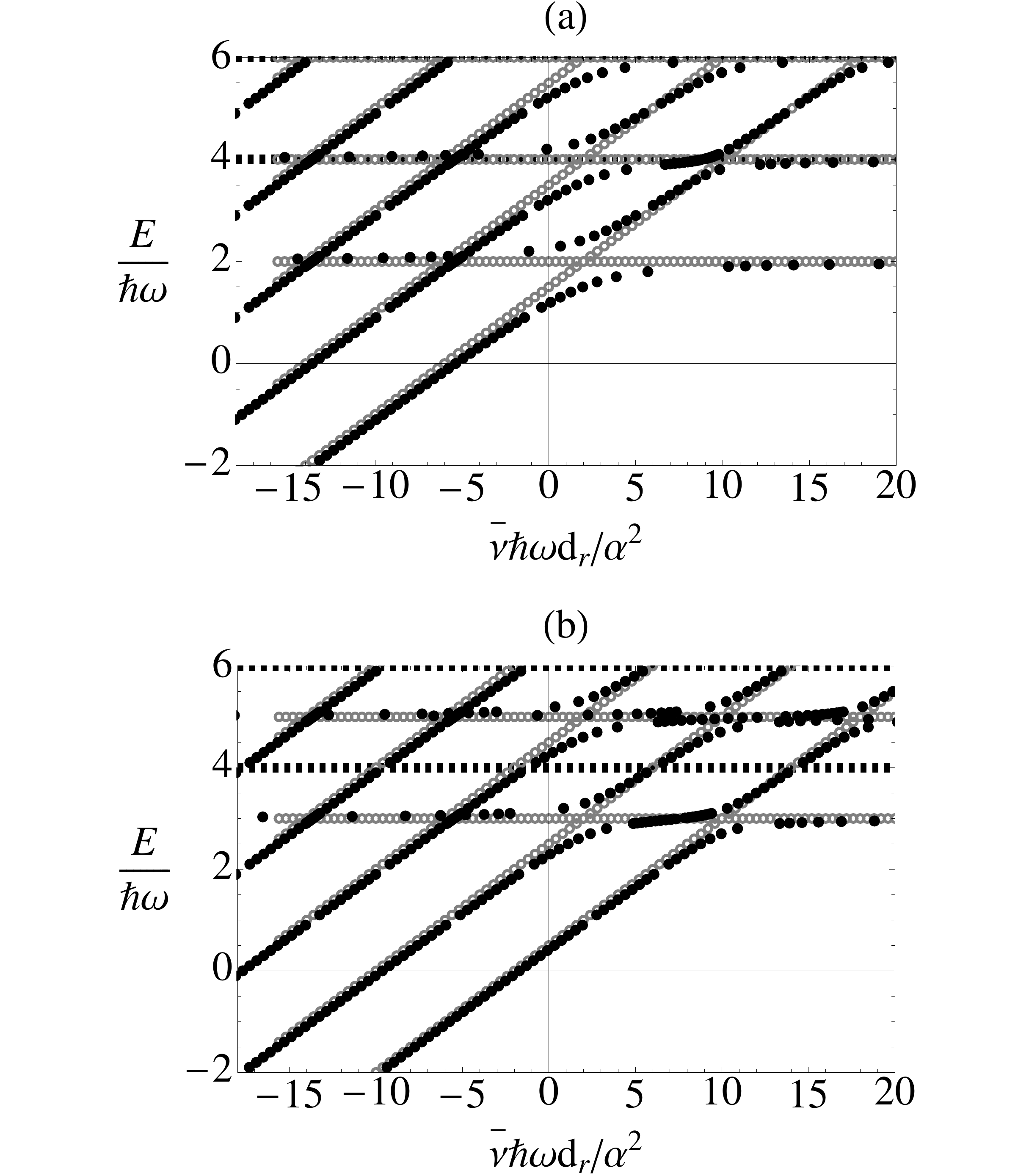}
\caption{Energy levels (in the centre-of-mass frame) versus detuning for three atoms in a harmonic well with $s$-wave Feshbach scattering at a dimensionless coupling $\tilde{\alpha}=0.5$. Dashed lines represent total pseudospin $S=3/2$ states which are unaffected by $s$-wave Feshbach scattering. Figures (a) and (b) represent eigenstates with odd and even inversion symmetry respectively. Grey circles represent eigenstates in the absence of Feshbach scattering. Black dots represent $S=1/2$ states.}
\label{Fig.0}
\end{figure}

We present in figure \ref{Fig.0} our results for the spectrum at
$\tilde{\alpha}=0.5$. We plot the energy in units of $\hbar\omega$
versus the dimensionless detuning parameter $\bar{\nu}\hbar\omega
d_r/\alpha^2$. The selection rules discussed above are reflected in
the energy spectrum as it can be seen in figure \ref{Fig.0}. The two
graphs, (a) and (b), correspond to the energy levels for states with
even and odd inversion symmetry.

The $S=3/2$ open-channel states which are unaffected by the Feshbach
scattering are found at energies $E/\hbar\omega = 4,6,\ldots$. At
negative detuning and at negative energies, we see the unperturbed
atom-dimer states $|\chi_n\rangle$, the different diagonal lines
corresponding to different atom-dimer oscillations. As we increase
$\bar\nu$, Feshbach interaction with open-channel states become
possible near the resonance points given by (\ref{eq:crossing}).  At
the resonances, there are anti-crossings between the three-atom and
atom-dimer states.  For small ${\tilde \alpha}$, we can approximate by
restricting the matrix equation (\ref{eq:coeffatomdimer}) to just the
two open and closed levels near the anti-crossing in question. Then,
we see immediately that, for resonances at positive detuning
$\bar\nu$, the size of the splitting at resonance is of the order of
the matrix element coupling the two states: $\alpha_{kln}\sim
\alpha/d_r^{1/2}=\tilde\alpha\hbar\omega$.

However, for resonances at negative $\bar\nu$, the splittings are much
smaller. As discussed in the section \ref{sec:selectionrules}, these
are the levels which are only coupled through multiple Feshbach
scattering.

\begin{figure}[hbt]
\includegraphics[width=\columnwidth]{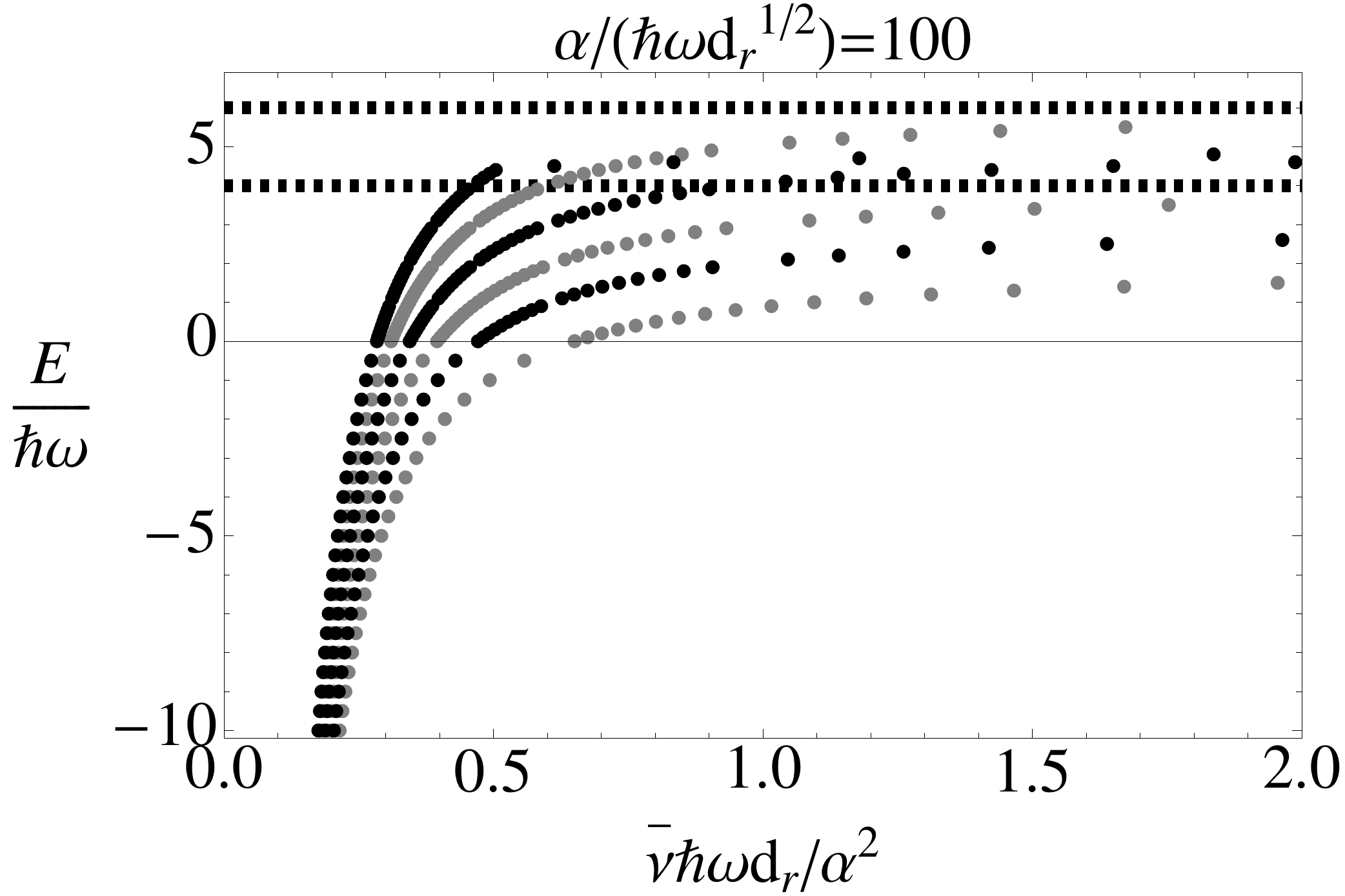}
\caption{Lowest six energy levels at strong Feshbach coupling:
$\tilde\alpha=100$. Dashed lines represent total pseudospin $S=3/2$
states. Grey and black dots represent $S=1/2$ states with odd and even
inversion symmetry respectively.}
\label{Fig.1}
\end{figure}

\begin{figure}[hbt]
\includegraphics[width=\columnwidth]{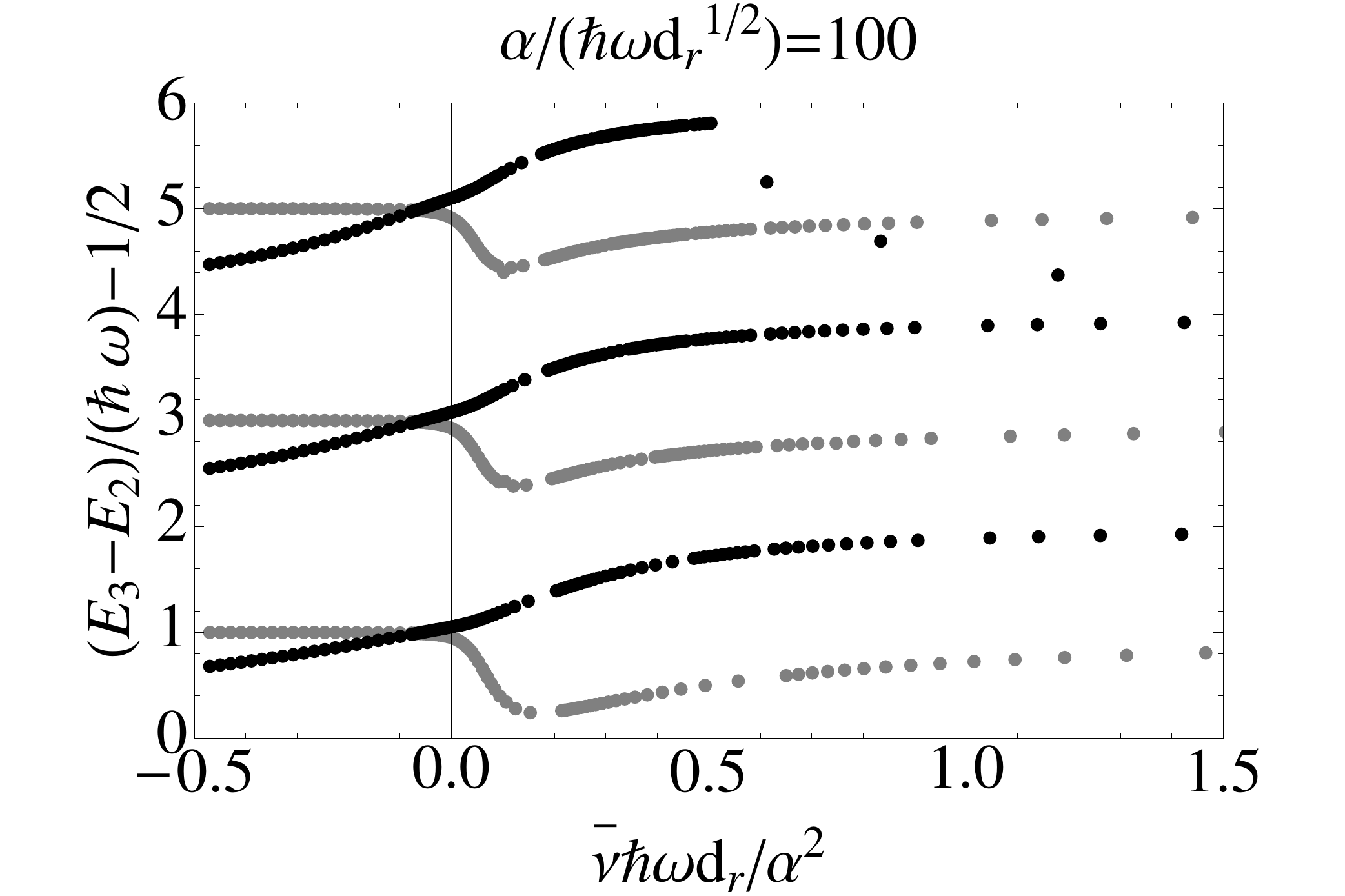}
\caption{Difference between the energy of three atoms in a harmonic
well and the ground-state energy of two atoms in the well plus the
lowest one-atom energy in a separate well. Grey and black dots
represent $S=1/2$ states with odd and even inversion symmetry
respectively.}
\label{Fig.2}
\end{figure}

We can now turn to the regime of strong Feshbach scattering. Large coupling strength $\tilde{\alpha}$ means much stronger level anticrossing than for $\tilde{\alpha}=0.5$ case shown in figure \ref{Fig.0}. As an example, we have calculated the spectrum for $\tilde{\alpha}=100$. (We will see later that there is very little admixture of high-energy 
level in the lowest-energy eigenstates, justifying our use of a relatively low cutoff $n_C$.)
The six levels with the lowest energies are shown in figure \ref{Fig.1}. 
The apparent kink in the highest branch in figures \ref{Fig.1} and \ref{Fig.2} is due to the level crossing
that can be better seen for the small $\tilde{\alpha}$ case in figure \ref{Fig.0}.
Again, we see the unperturbed open-channel states at $4\hbar\omega$ and $6\hbar\omega$ (dashed lines).
Note that the lines corresponding to the unperturbed atom-dimer states would be very steep and very close to the vertical axis on this graph.

We find level crossings in the low-energy spectrum between these two
sets of states with opposite parity under inversion. This is most
easily seen in energy spectrum at weak Feshbach scattering
(Fig.~\ref{Fig.0}). For large positive detuning, the lowest energy
states belong to the odd-parity branch that gives the open-channel
$|01\rangle$ state asymptotically far away from the resonance
($\bar\nu\rightarrow +\infty$). On the other hand, for large negative
detuning, the states of lowest energy belong to the even-parity branch
that becomes the $|\chi_0\rangle$ atom-dimer state asymptotically
($\bar\nu\rightarrow -\infty$). Therefore, the two parity branches
must cross as a function of detuning.  For weak Feshbach scattering,
the crossing for the lowest two parity states occurs at a positive
detuning, just below $E = 2\hbar\omega$.  For strong Feshbach
scattering, we find that these level crossings have shifted to small
negative detuning and negative energies ($\bar\nu \hbar\omega
d_r/\alpha^2 \simeq -0.07$ and $E\simeq -9\alpha/d_r^{1/2}$ for
$\tilde\alpha=100$). Similar features were found for the spectrum of
three atoms in a 3D harmonic well within the one-channel model
\cite{Kestner}. 

The level crossings between the branches of opposite inversion
symmetry are best seen if we subtract from the three-body energies
($E_3$) both the ground-state energy of two atoms in a well ($E_2$ as
given by (\ref{eq:twobodyenergy})) and the energy of one atom in a
separate well ($\hbar\omega/2$). This quantity, $E_3 - E_2 -
\hbar\omega/2$, is plotted in figure \ref{Fig.2}. The subtraction does
not alter the crossings of the three-body states at any given
$\bar\nu$. In section \ref{sec:summary}, we speculate that these
crossings may lead to quantum phase transitions.

This data allows us to discuss the ground state of the system with
three atoms for every two wells.  One might ask whether the Feshbach
interaction could generate an attraction such that there is a regime
where the system would prefer to arrange three atoms in one well and
none in the other well.  Such a distribution would have an energy of
$E_3$ plus $\hbar\omega/2$ for the zero-point motion of the centre of
mass. We can also have two atoms in one well and one atom in the
ground state in the other well.  This has energy $E_2$ plus
$\hbar\omega/2$ for the centre of mass in one well and $\hbar\omega/2$
in the other well. Therefore, the energy gap between these two states
is $E_3 - E_2 - \hbar\omega/2$. From Fig.~\ref{Fig.2}, we see that
this gap remains positive for all detunings, suggesting that the
ground state of an optical lattice with 3/2 atoms per site to be a
phase with uniform density. This energy gap has a minimum as small as
0.2$\hbar\omega$ at $\tilde\alpha=100$. This minimum is due to a
reduction in $E_3$ compared to $E_2$ arising from exchange processes
which do not occur for two atoms, as discussed in previous section
\ref{sec:model} after equation (\ref{eq:ME4}). We note that this
energy gap is much smaller in magnitude than the Feshbach energy scale
$\alpha/d^{1/2}_r$ and the individual energies $E_3$ and $E_2$.

We can similarly ask about the ground state of four atoms in two
wells. The evenly-distributed state with two atoms per well is lower
in energy because it contains two binding energies $2E_2$ whereas the
uneven distribution can only take advantage of Feshbach physics for
one pair of atoms in $E_3$.
\begin{figure}[hbt]
\includegraphics[width=\columnwidth]{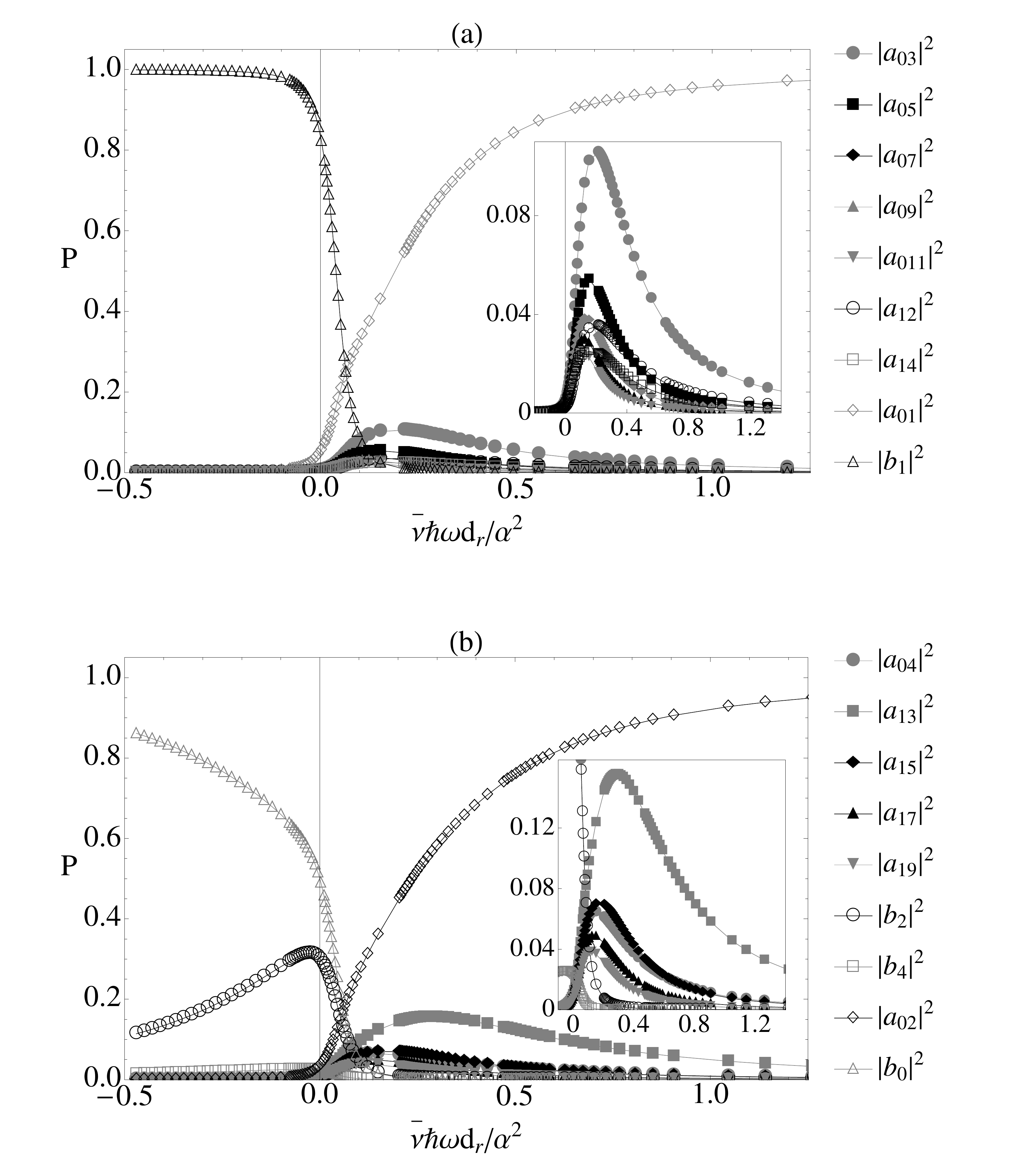}
\caption{Ground and first excited pseudospin $S=1/2$ state composition
as a function of detuning $\bar\nu$ ($\tilde\alpha=100$).  The
probabilities $|a_{kl}|^2$ and $|b_n|^2$ correspond to open-channel
and atom-dimer probabilities respectively (see the equation
(\ref{eq:eigenstate}) for definition of $a_{kl}$ and $b_n$). (a)
Lowest-energy state with odd inversion symmetry (lowest gray
data in figure \ref{Fig.1}). (b) Lowest-energy state with even
inversion symmetry (lowest black data in figure
\ref{Fig.1}).}
\label{Fig.3}
\end{figure}

Having discussed the energy spectrum of the three-body system, we will
now examine the wavefunctions of the low-energy eigenstates.  The
compositions of the ground state and the first excited state (for
pseudospin $S$=1/2) are shown at figure \ref{Fig.3}. Figure
\ref{Fig.3} (a)/(b) presents the composition of the lowest energy
state with an odd/even inversion symmetry.  As expected, the state in
figure (a) becomes the $|\chi_1\rangle$ atom-dimer level at large
negative detuning while it is predominately the open-channel
$|0,1\rangle$ level at large positive detuning.  Near the resonance,
there is also a significant admixture of other open-channel levels and
a smaller admixture of other atom-dimer levels. The most significant
components are shown at the figure. 
Figure (b) gives an example with even inversion
symmetry. The state is asymptotically the $|\chi_0\rangle$ atom-dimer
state at large negative detunings. Closer to the resonance at negative
$\bar{\nu}$ there is also a significant $|\chi_2\rangle$
component. For positive $\bar{\nu}$, the dominant contributions are
from the $|0,2\rangle$ and $|1,3\rangle$ open-channel levels with the
$|1,3\rangle$ contribution vanishing far from the resonance.

We notice that, while only $|\chi_1\rangle$ dominates in the lowest
energy state with odd inversion symmetry, both $ |\chi_0\rangle$ and
$|\chi_2\rangle$ are significant for the corresponding state with even
inversion symmetry particularly near the resonance. This difference
can be traced back to the Feshbach matrix element $\alpha_{kln}$.  As
discussed in \ref{sec:selectionrules}, the wavefunctions involved
should have similar symmetries to ensure a finite value for
$\alpha_{kln}$. Thus, using the rules that $k \neq l$ (Pauli
principle) and that $k+l$ and $n$ should be both even or both odd with
$n\leq k+l$, the lowest few $\alpha_{kln}$ that contribute to the
lowest energy states are $\alpha_{011}$ and $\alpha_{013}$ for the
odd-parity state and $\alpha_{020}$ and $\alpha_{022}$ for the
even-parity state.  These matrix elements are large only if the
quantum numbers $k$ and $l$ are similar to $n$ and so: $|\alpha_{011}|
\gg |\alpha_{013}|$ and $|\alpha_{020}| \sim |\alpha_{022}|$. In terms
of equation (\ref{eq:coeffatomdimer}) for the coupling of the
closed-channel amplitudes, the magnitude of $M_{11}$ is big compared
to that of $M_{13}$ for the odd case, while both $M_{00}$ and $M_{02}$
are significant for the even case. Hence, both $ |\chi_0\rangle$ and
$|\chi_2\rangle$ contribute to the lowest energy even sector
eigenstate in Fig.~\ref{Fig.3}(b).

These results allow us to understand the validity of our truncation of
the matrix equation (\ref{eq:ME1}). We see that the admixture of
high-energy levels is small. This allows us to ignore the contribution
of $|\chi_n\rangle$ levels with $n\geq 2n_C = 40$. For comparison, if
we use a smaller cutoff of $n_C=5$ instead of 20, the value of $\bar\nu$
at a given $E$ changes by less than 1\% and the ratios between different 
amplitudes change by less than 10\%.

\section{Comparison with one-channel model}
\label{sec:onechannel}

In this paper, we have used a two-channel model for states near the
Feshbach resonance. It is often convenient to use a one-channel model
in which the Feshbach-induced interaction between open-channel states
is modelled by a contact potential: $V(\vec{r})=g\delta(\vec{r})$,
with a potential strength $g$ that depends on the
detuning\cite{Busch,Kestner}. We discuss now how a one-channel
model may be appropriate for the three-body problem in a 1D well.

Physically, we expect that an effective model for just the open
channel cannot capture physics involving the details of the dimer
state in the closed channel. Therefore, one-channel models can
approximate the results of two-channel models only if we focus our
attention on scattering states and \emph{weakly} bound states, neither
of which would have a strong admixture of the closed-channel dimer. We
have been discussing the branch of bound states induced by Feshbach
interaction. In three dimensions without a harmonic trap, these states
are weakly bound, with a binding energy that vansihes as we approach
the resonance ($\bar\nu=0$) from negative detuning. In other words,
these eigenstates have only a small amplitude in the closed channel
dimer state. Therefore, the matching to a one-channel model is best
\emph{near} the resonance. The range of validity of the one-channel
model will be wide if the Feshbach coupling is strong. 

This comparison between one-channel and two-channel models in 3D has
also been studied in the presence of a harmonic trap.  At a wide
Feshbach resonance (strong coupling between the open and closed
channels), the admixture of the closed channel dimer component to the
low energy states is very small \cite{Diener} and the effective
interaction between the atoms close to the resonance can be described
by an effective contact potential \cite{Busch,Kestner,Diener2,Diener3}
with $g_{3D}=2\pi\hbar^2a_{\rm 3D}/\mu$ which diverges as one sweeps
the system across the resonance at $\bar\nu=0$. However, there are
also narrow resonances where a one-channel model does not apply
\cite{Tiesinga} due to the significant admixture of the closed-channel
component.

We will now turn to our results in one dimension.  We have already
studied in section \ref{sec:twobody} the matching of the two-channel
results (for two atoms without a trap) to one-channel scattering
parameters for the branch of bound states that develops at energies
below the open channel. Note that this branch is not confined to
$\bar\nu<0$ (as is the case in 3D without a trap). Instead, it exists
for all $\bar\nu$ with the most weakly bound states at large positive
detuning. This is the regime where we expect the one-channel and
two-channel models to converge.  The weak-binding condition that $|E|
\sim \hbar^2/2\mu a^2_{\rm 1D} \ll \bar\nu$ can be expressed in terms
of the dimensionless Feshbach and detuning parameters used in our
study (\ref{eq:scatteringparms}):
\begin{equation}\label{eq:matchinglimit1D}
\frac{\bar\nu\hbar\omega d_r}{\alpha^2} \simeq 
\frac{a_{\rm 1D}}{d_r} \gg \frac{1}{\tilde\alpha^{2/3}}.
\end{equation}
We see that this matching applies to positive detuning \emph{away}
from the level anticrossing near $\bar\nu=0$. Note that the regime of
validity improves at strong coupling: $\tilde\alpha \rightarrow
\infty$ so that such a theory may be applicable for realistic $^{40}$K
systems except very close to the resonance.  In terms of an effective
contact potential strength, we have $g_{\rm
1D}=-\hbar^2/\mu a_{\rm 1D}$ \cite{Busch} so that the regime of
validity corresponds to a weak contact potential which traps a bound state for
arbitrary weak $g_{\rm 1D}$ in one dimension.

These conclusions about the validity of a one-channel approximation
are also consistent with our three-atom results presented above. 
We can examine directly the degree of the admixture of open- and closed-channel components as a
function of detuning. At large negative detuning, the states
illustrated in figure \ref{Fig.3} are predominantly the closed-channel
atom-dimer state and so should have no relation to any model based on
open-channel states. The change in the balance between open- and
closed-channel components as a function of increasing detuning is
clearly seen in figure \ref{Fig.3}. Indeed, in our 1D system, there
is a significant admixture of the atom-dimer levels close to the
resonance even for strong coupling between the open and closed
channels ($\tilde{\alpha}\gg 1$). Therefore, we conclude that the
system cannot be modelled by a one-channel theory close to the
resonance.  On the other hand, for large positive detuning, the states
are predominantly open-channel states and so can be modelled in a
effective theory for the open-channel states.

\begin{figure}[hbt]
\includegraphics[width=\columnwidth]{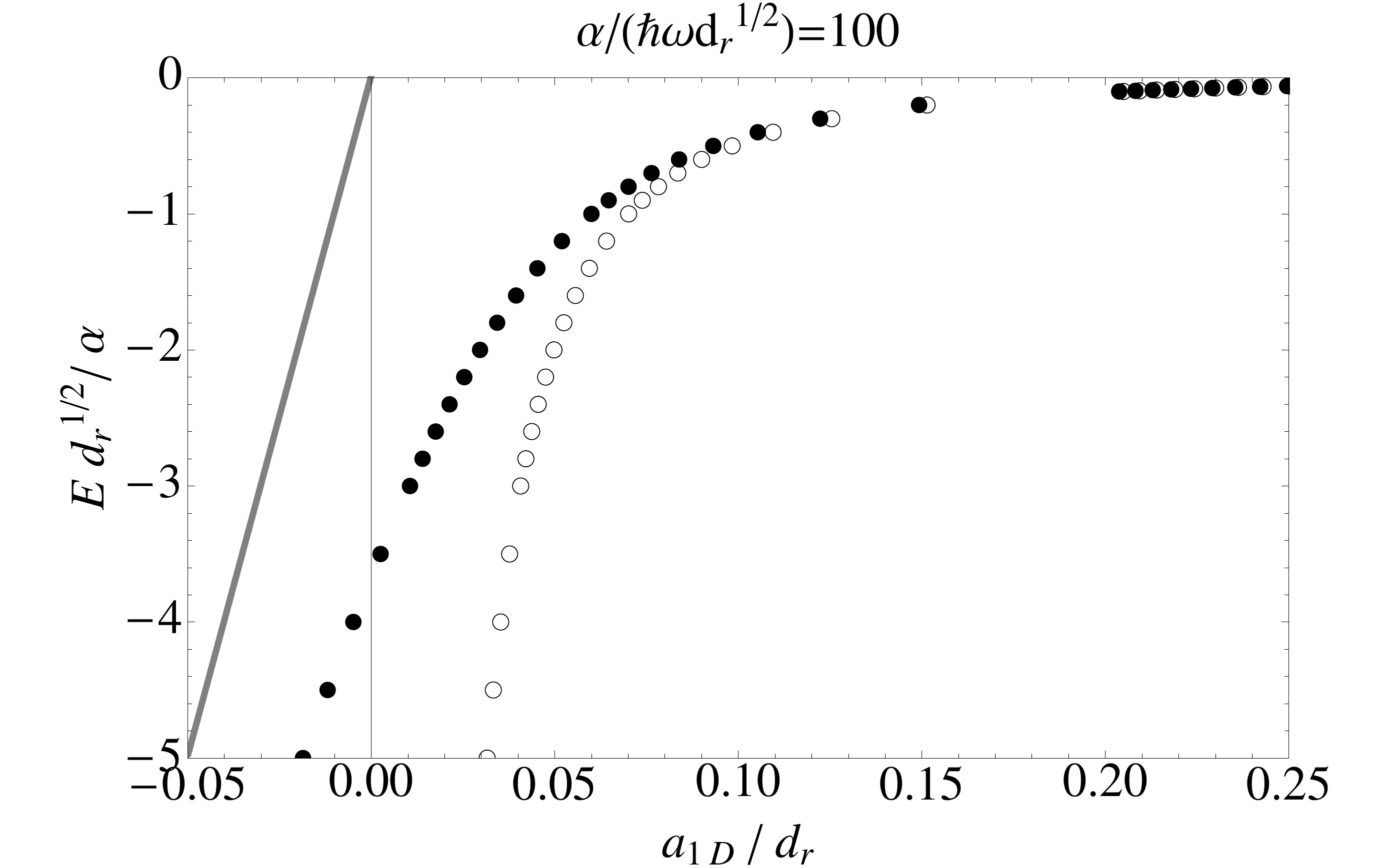}
\caption{The lowest energy state with odd inversion symmetry for three atoms
  calculated using the two-channel models (solid circles) and
  one-channel model (open circles). The horizontal axis is plotted using the 
correspondence in equation (\ref{eq:matchinglimit1D}). Grey line represents
the closed-channel state in the two-channel model in the absence of Feshbach coupling 
($E=\bar\nu+3\hbar\omega/2$).}
\label{fig:onetwochannel}
\end{figure}
Furthermore, we have performed a one-channel calculation for three
atoms in a well for our system using an effective $g_{\rm 1D}$
\cite{Thesis}, following an analogous study for the 3D problem
\cite{Kestner}. From Fig.~\ref{fig:onetwochannel}, we see that the
energies calculated for the lowest branch of eigenstates agree in the
one-channel and two-channel models at large positive detuning, but
they differ significantly close to the resonance.  Similar conclusions
can be made for the two atoms in a 1D well by comparing one-channel
results \cite{Busch} the two-channel result (\ref{eq:twobodyenergy})
\cite{Diener}.  We can also confirm that the one- and two-channel
calculations give different results when there is a significant
admixture of the atom-dimer levels.

Interestingly, we find that the energy \emph{difference} $E_3-E_2$ are
very similar in the one-channel and two-channel models for the whole
range of positive detuning at $\tilde\alpha=100$.

\section{Summary and Discussion}
\label{sec:summary}
We have considered two species of fermionic $^{40}$K atoms at ultra
cold temperatures trapped in a 1D optical lattice and close to an
inter-species Feshbach resonance. We have concentrated on the
zero-tunnelling limit in which the dynamics of the system is well
described by the local spectrum of atoms at a lattice site. The
calculations are performed within the two-channel model for two and
three atoms at a lattice site approximated by a harmonic
well. 

Close to the resonance, we find a significant admixture of the
atom-dimer states to the lowest energy eigenstates of the
system. Consequently, the one- and two- channel models must differ
close to the resonance for our 1D system since a one-channel theory
cannot capture the physics of the dimer state.

The results suggest that the ground state is a state with uniform atom
density.  In particular, for three atoms in two wells, we compared the
energy of all three atoms in one well (3+0) to the energy of two atoms
in one well and one atom in the other (2+1). The energy difference can
be as small as 0.2$\hbar\omega$ at $\tilde\alpha=100$ which is much
smaller than the Feshbach energy scale. However, we note that, for the
$S_z=+1/2$ case, we have neglected the residual interaction between
the atom and dimer in a $|\chi_n\rangle$ state. There is an attractive
component from the closed-channel potential $V^{\rm cl}$ and also a
Feshbach scattering term, both of which we argued to be suppressed by
fermionic statistics. We can ask whether these additional terms may
significantly reduce or even change the sign of the energy difference
between 2+1 and 3+0 configurations. In other words, we ask whether
there is an instability to a density wave in an optical lattice with a
filling close to 3/2 atoms per site.

To obtain a quantitative analysis for such a scenario, we would
require a detailed short-distance description of the dimer
state. Here, we will limit ourselves to a rough estimate for the
lowest-energy state at positive detuning by considering the
probability that the spectator atom can be in the vicinity of the
dimer.

This state has odd inversion symmetry and the amplitude for the two
atoms ($r_1$ and $r_2$) to approach each other is proportional to
$(r_1-r_2)/d_r^{1/2}$. So, the probability for the spectator to be in
the vicinity of a dimer of size $d_m$ is proportional to $\int^{d_m}
(r/d_r)^2 dr/ d_r \sim (d_m/d_r)^3$.  We estimate the residual
attraction $\epsilon_{\rm ad}$ between the atom and the dimer would be
of the order of $\bar V (d_m/d_r)^3$ where $\bar V$ is a measure of
the strength $V^{\rm cl}$ experienced by the closed-channel dimer,
\emph{e.g.}~the minimum value of the potential. As mentioned above,
the minimum energy gap in $E_3-E_2-\hbar\omega/2$ is only a fraction
of $\hbar\omega$. Therefore, if $\bar V$ is sufficiently large
compared to $\hbar\omega$, there is a possibility that this additional
attraction can favour the 3+0 configuration compared to the 2+1
configuration in two wells. (A similar term in the Feshbach scattering
can be absorbed into a renormalised scattering strength $\alpha$, but our results are
not sensitive to the value of $\alpha$ in this regime of large $\tilde\alpha$.)
In fact, the even-symmetry state has a similar but larger correction of $\bar V d_m/d_r$
which could change the ordering in energy of the even and odd branches over a range
of detuning. These speculations need to be confirmed by further work.

Note that these hypothetical scenarios rely on the residual interaction between
an $\uparrow$ atom (but not a $\downarrow$ atom) with a dimer
in the state $|\uparrow\circ\rangle$.  In other words, such a scenario would
reveal the asymmetry of the system under the transformation
$\uparrow\leftrightarrow\downarrow$, it would only be a possibility if
we have more atoms in the $\uparrow$ open-channel state than in the
$\downarrow$ state.  The magnitude of this effect depends on the ratio
of the size of the dimer in the closed channel compared to the trap
size and therefore depends on the details of the actual Feshbach
resonance.

Finally, we discuss briefly the optical lattice with weak tunnelling
between a set of wells. Consider the case with three atoms per site on
average. From the level crossing we found in figure \ref{Fig.2}, we
see that the ground state would consist of states with odd or even
inversion symmetry within each site depending on the detuning
parameter. As we sweep past this crossing, this may be a first-order
transition. However, it is also possible that a continuous quantum
phase transition would occur because an inter-site tunnelling would
couple the states of different inversion symmetries. We need to
investigate collective quantum fluctuations to study this possibility.

\end{document}